\documentclass[superscriptaddress,twocolumn,showpacs,pre,floatfix]{revtex4}

\usepackage{amsmath}
\usepackage{color}
\usepackage{tabularx}
\usepackage[dvips]{graphicx}

\usepackage[greek,english]{babel}
\usepackage{teubner}

\def\koppa{\hbox{\foreignlanguage{greek}{\coppa}}}

\begin{document}

\title{Hyperscaling breakdown and Ising Spin Glasses: the Binder cumulant}

\author{P. H.~Lundow} \affiliation{Department of Mathematics and
  Mathematical Statistics, Ume{\aa} University, SE-901 87 Ume{\aa},
  Sweden}

\author{I. A.~Campbell} \affiliation{Laboratoire Charles Coulomb
  (L2C), Univ. Montpellier, CNRS, Montpellier, France.}

\begin{abstract}
  Among the Renormalization Group Theory scaling rules relating
  critical exponents, there are hyperscaling rules involving the
  dimension of the system.  It is well known that in Ising models
  hyperscaling breaks down above the upper critical dimension. It was
  shown by M. Schwartz [Europhys. Lett. {\bf 15}, 777 (1991)] that the
  standard Josephson hyperscaling rule can also break down in Ising
  systems with quenched random interactions. A related Renormalization
  Group Theory hyperscaling rule links the critical exponents for the
  normalized Binder cumulant and the correlation length in the
  thermodynamic limit. An appropriate scaling approach for analyzing
  measurements from criticality to infinite temperature is first
  outlined. Numerical data on the scaling of the normalized
  correlation length and the normalized Binder cumulant are shown for
  the canonical Ising ferromagnet model in dimension three where
  hyperscaling holds, for the Ising ferromagnet in dimension five (so
  above the upper critical dimension) where hyperscaling breaks down,
  and then for Ising spin glass models in dimension three where the
  quenched interactions are random.  For the Ising spin glasses there
  is a breakdown of the normalized Binder cumulant hyperscaling
  relation in the thermodynamic limit regime, with a return to size
  independent Binder cumulant values in the finite-size scaling regime
  around the critical region.
\end{abstract}

\pacs{75.50.Lk, 75.40.Mg, 05.50.+q}

\maketitle

\section{Introduction}\label{sec:I}
The consequences of the Renormalization Group Theory (RGT) approach
have been studied in exquisite detail in numerous regular physical
models, typified by the canonical near-neighbor interaction
ferromagnetic Ising models. It has been tacitly assumed that
Edwards-Anderson Ising Spin Glasses (ISGs), where the quenched
interactions are random, follow the same basic scaling and
Universality rules as the Ising models.

The Binder cumulant \cite{binder:81} is an important observable which
has been almost exclusively exploited numerically for its scaling
properties as a dimensionless observable very close to criticality in
the finite-size scaling (FSS) regime $L \ll \xi(\beta)$, where $L$ is
the sample size and $\xi(\beta)$ is the second-moment correlation length
at inverse temperature $\beta$. Here we will consider its scaling
properties over the whole temperature region, in particular in the
Thermodynamic limit (ThL) regime $L \gg \xi(\beta)$ where the
properties of a finite-size sample normalized appropriately are
independent of $L$ and so are the same as those of the infinite-size
model.

We will explain in detail the overall scaling analysis procedure,
based on Ref.~\cite{wegner:72,privman:91,campbell:06}, which we use in
both the cases of standard Ising models and of ISGs.

\section{Scaling}\label{sec:II}
In numerical simulation analyses the conventional RGT based approach
consists in using as the thermal scaling variable the reduced
temperature $t = (T-T_{c})/T_{c}$, together with the principal
observables $\chi(t,L)$ the susceptibility, $\xi(t,L)$ the second
moment correlation length, and $g(t,L)$ the Binder cumulant. (For
finite-size simulation data the standard finite-$L$ definition for the
second moment correlation length $\xi(\beta,L)$ through the Fourier
transformation of the correlation function is used, see for instance
Ref.~\cite{hasenbusch:08} Eq. $14$).  The conventional approach is
tailored to the critical region; however at high temperatures $t$
diverges and $\xi(t,L)$ tends to zero, so it is not possible to
analyse the entire paramagnetic regime without introducing diverging
correction terms. For the Ising systems this problem can be eliminated
by using the inverse temperature $\beta=1/T$, a practice which
pre-dates RGT.

The thermal scaling variable $t$ is also widely used in analyses of
simulation data in ISGs. As the relevant interaction strength in ISGs
is $[\langle J_{ij}^2\rangle]$, the symmetric interaction distribution
ISG thermal scaling variable should logically depend on the square of
the temperature; this basic point was made some thirty years ago
\cite{singh:86} but has generally been ignored.

As a basis for a rational scaling approach which englobes the entire
paramagnetic region so including both the finite-size scaling regime
(FSS, $L \ll \xi(\beta,\infty)$) and the thermodynamic-limit regime
(ThL, $L \gg \xi(\beta,\infty)$), we start from the Wegner ThL scaling
expression for the Ising susceptibility \cite {wegner:72}
\begin{equation}
  \chi(\tau)= C_{\chi}\tau^{-\gamma}\left(1+ a_{\chi}\tau^{\theta} +
    b_{\chi}\tau + \cdots\right)
\end{equation}
where $\tau = 1-\beta/\beta_{c}$ with $\beta$ the inverse temperature.
(The Wegner expression is often mis-quoted with $t$ replacing
$\tau$). The terms inside $(..)$ are scaling corrections, with
$\theta$ the leading correction exponent which is universal for all
observables. As $\tau$ and $\chi(\tau)$ both tend to $1$ at infinite
temperature, the whole paramagnetic region can be covered without
divergencies, to good precision when a small number of well-behaved
correction terms are included. (To obtain infinite precision an
infinite number of correction terms would be needed, just as in
standard FSS analyses perfect precision in principle requires a series
of corrections to infinite $L$). In ISG models where the interaction
distributions are symmetric about zero, an appropriate thermal scaling
variable to be used with the same Wegner expression is $\tau =
1-(\beta/\beta_{c})^2$,
Refs.~\cite{singh:86,klein:91,daboul:04,campbell:06}.  In the ThL
regime $L \gg \xi(\beta)$ the properties of a finite-size sample, if
normalized correctly, are independent of $L$ and so are the same as
those of the infinite-size model. A standard rule of thumb for the
approximate onset of the ThL regime is $L > 7\xi(\beta,L)$ and the
ThL regime can be easily identified in simulation data.  An important
virtue of this approach is that the ThL numerical data can be readily
dovetailed into High Temperature Series Expansion (HTSE) values
calculated from sums of exact series terms (limited in practice to a
finite number of terms). No such link can be readily made when the
conventional FSS thermal scaling variable $t$ is used.

To apply the Wegner formalism to observables $Q$ other than $\chi$, we
introduce the rule that these observables should be normalized in such
a way that the infinite-temperature limit $Q(\tau=1) \equiv
1$, without the critical limit being modified. For the susceptibility
with the standard definition no normalization is required as this
condition is automatically fulfilled, with a temperature-dependent
effective exponent $\gamma(\tau) =
\partial\ln\chi(\tau,L)/\partial\ln\tau$ in Ising models and in
ISGs with the appropriate $\tau$. Then
\begin{equation}
  \gamma(\tau)=\gamma -
  \frac{a_{\chi}\theta\tau^{\theta}+b_{\chi}\tau}{1 +
    a_{\chi}\tau^{\theta}+b_{\chi}\tau}
\end{equation}
to second order in the corrections \cite{lundow:15}.

In Ref.~\cite{campbell:06} the normalized second-moment correlation
length was introduced : $\xi(\tau,L)/\beta^{1/2}$ in Ising models and
$\xi(\tau,L)/\beta$ in ISG models. From exact and general HTSE
infinite-temperature limits, this normalized correlation length tends
to exactly $1$ at infinite temperature \cite{butera:02,daboul:04}. The
temperature-dependent effective exponent is $\nu(\tau) =
\partial\ln[\xi(\tau,L)/\beta^{1/2}]/\partial\ln\tau$ in Ising
models and $\nu(\tau) =
\partial\ln[\xi(\tau,L)/\beta]/\partial\ln\tau$ in ISG models. A
Wegner-like relation is
\begin{equation}
  \xi(\tau)/\beta^{1/2}= C_{\xi}\tau^{-\nu}\left(1+
  a_{\xi}\tau^{\theta} + b_{\xi}\tau + \cdots\right)
\end{equation}
so
\begin{equation}
  \nu(\tau)= \nu -\frac{a_{\xi}\theta\tau^{\theta}+b_{\xi}\tau}{1
    + a_{\xi}\tau^{\theta}+b_{\xi}\tau}
\end{equation}
The critical limiting ThL exponent $\nu$ is unaltered by this
normalization (models with zero critical temperatures are a special
case). The normalized correlation length can be accurately expressed
over the entire paramagnetic region with a limited number of generally
weak correction terms. The temperature-dependent effective exponents
$\gamma(\tau)$ and $\nu(\tau)$ are well-behaved over the whole
paramagnetic regime with the exact infinite-temperature hypercubic
lattice limits for Ising models of $\gamma(1) = 2D\beta_{c}$ and
$\nu(1) = D\beta_{c}$, and for the ISG models $\gamma(1) =
2D\beta_{c}^2$ and $\nu(1) = (D-K/3)\beta_{c}^2$ where $K$ is the
kurtosis of the interaction distribution and $D$ is the dimension.
The normalized Binder cumulant scaling is discussed below.

\section{Hyperscaling}\label{sec:III}
Among the standard rules linking critical exponents are the
hyperscaling relations \cite{widom:65,kadanoff:66,josephson:66}.  A
textbook definition of hyperscaling is : "Identities obtained from the
generalised homogeneity assumption involve the space dimension D, and
are known as hyperscaling relations."~\cite{simons:97}. The most
familiar form of the hyperscaling relation is $\alpha = 2 -D\nu$ which
through the Essam-Fisher relation $\alpha + D\nu -2\Delta=2$ can be
re-written $2\Delta = \gamma +D\nu$.  $ \Delta$ is the Gap exponent,
defined \cite{butera:02} through the critical behavior of the higher
field derivatives of the free energy, $\gamma_{k} = \gamma
+(k-2)\Delta$ ; $\Delta = \gamma +\beta$ \footnote{Explicitly quoting
  Ref.~\cite{pelissetto:02} : "Below the upper critical dimension, the
  following hyperscaling relations are supposed to be valid:
  $2-\alpha = D\nu$,
  $2\Delta_{\mathrm{gap}} = D\nu +\gamma$
  where $\Delta_{\mathrm{gap}}$ is the gap exponent, which controls
  the radius of the disk in the complex-temperature plane without
  zeroes, i.e. the gap, of the partition function (Yang-Lee
  theorem)".}.

This form of the hyperscaling relation has practical consequences for
the scaling of the normalized Binder cumulant.  Hyperscaling is well
established in standard models, such as the Ising models in dimensions
less than the upper critical dimension, see Section~\ref{sec:IV}. The
specific case of breakdown of hyperscaling for the Ising model in
dimension $5$, above the upper critical dimension, is discussed in
Section~\ref{sec:V}.

In Ising ferromagnets, in the thermodynamic limit (infinite size
or $L \gg \xi(\tau)$) regime the susceptibility $\chi(\beta)$ scales
with the critical exponent $\gamma$, and assuming hyperscaling the
critical exponent for the second field derivative of the
susceptibility $\chi_{4}(\beta)$ (also called the non-linear
susceptibility) is ~\cite{butera:02}
\begin{equation}
  \gamma_{4}=\gamma +2\Delta_{\mathrm{gap}}  = D\nu + 2\gamma
  \label{gam4}
\end{equation}

Note that $\chi_{4}$ in a hypercubic lattice is directly related to
the Binder cumulant through
\begin{equation}
  2g(\beta,L) = \frac{-\chi_{4}}{L^D\chi^{2}} = \frac{3\langle
    m^2\rangle^2 - \langle m^4\rangle}{\langle m^2\rangle^2}
\end{equation}
see Eq.~(10.2) of Ref.~\cite{privman:91}.  Thus in the ThL regime the
normalized Binder cumulant $L^D g(\beta,L)$ (or alternatively
$-\chi_{4}(\beta)/(2\chi(\beta)^2)$) scales with the critical exponent
$(\nu D + 2\gamma) - 2\gamma = \nu D$, together with correction terms
as for any such observable, because of the RGT scaling and
hyperscaling~\cite{widom:65,josephson:66} relationships between exponents. Here
$\nu$ and $\gamma$ are the standard critical exponents for the
correlation length and the susceptibility.

It can be noted that in any $S = 1/2$ Ising system the
infinite-temperature (i.e. independent spins) limit for the Binder
cumulant is $g(0,N) \equiv 1/N$, where N is the number of spins; as $N
= L^{D}$ for a hypercubic lattice, at infinite temperature
$L^{D}g(\tau,L) \equiv 1$. Thus the Ising normalized Binder cumulant
obeys the high temperature limit rule for normalized observables
introduced above.

Two forms of the hyperscaling
relation have been quoted above; the first is well known and concerns
the specific heat exponent $\alpha$. Many years ago this first
hyperscaling relation was predicted by Schwartz to break down in
quenched random systems \cite{schwartz:91}.  The breakdown of this
hyperscaling relation in the Random Field Ising model (RFI) has been
extensively studied \cite{gofman:93,vink:10,fytas:13}. Ising spin
glasses (ISGs) are also systems with quenched randomness in which
hyperscaling might be expected to break down by a generalisation of
Schwartz's argument. The exponent $\alpha$ in ISGs is always strongly
negative and so is very hard to measure directly; we will explore only
the second form of the hyperscaling relation which is less well known. We are
aware of no tests of this hyperscaling relation in ISGs.

We observe that in ISGs the ThL susceptibility $\chi(\tau)$ and the
normalized second moment correlation length $\xi(\tau)T$ follow the
Wegner scaling rules with only weak corrections over the entire
temperature range from criticality to infinity as has already been
shown, e.g. Ref.~\cite{campbell:06,lundow:15a}.  However, in ISGs for the
normalized Binder cumulant $L^{D}g(\tau,L)$ we indeed find very strong
deviations from the behaviour expected if hyperscaling held. Because
of difficulties inherent to ISG simulations, the ThL temperature range
attainable in the present measurements is restricted so these
deviations cannot be fully characterized, though it seems unlikely
that a huge and unspecified \lq\lq correction term\rq\rq{} should just
appear by accident.

\section {Privman-Fisher scaling and Extended scaling}\label{sec:IV}
Up to now we have been considering only data in the ThL.  Writing for
an Ising ferromagnet model with the normalized correlation length
$\xi(\tau,\infty) \sim \beta^{1/2}\tau^{-\nu}$ in the ThL and for an
observable $Q(\tau,L)$ where $Q(\tau,\infty) \sim \tau^{-q\nu}$ at
criticality, ignoring corrections to scaling the Privman-Fisher
finite-size rule \cite{privman:91} can be written

\begin{equation}
  \frac{Q(\tau,L)}{Q(\tau,\infty)} = F[L/\xi(\tau,\infty)] =
  F[LT^{1/2}\tau^{\nu}] 
\end{equation}
or
\begin{equation}
  Q(\tau,L) =
  F[LT^{1/2}\tau^{\nu}]/\tau^{q\nu}
\end{equation}
or
\begin{equation}
  \frac{Q(\tau,L)}{(LT^{1/2})^{q}} =
  \frac{F[LT^{1/2}\tau^{\nu}]}{\tau^{q\nu}(LT^{1/2})^{q}}
  = \frac{F[LT^{1/2}\tau^{\nu}]}{(\tau^{\nu}LT^{1/2})^{q}}
\end{equation}
or finally 
\begin{equation}
  \frac{Q(\tau,L)}{(LT^{1/2})^{q}} = F^{*}[(LT^{1/2})^{1/\nu}\tau]
\end{equation}
which is the "extended scaling" form of Ref.~\cite{campbell:06}. For
an ISG, $T^{1/2}$ is replaced throughout by $T$. If Wegner corrections
to scaling factors have been measured from ThL data, these can be
readily introduced into either the Privman-Fisher expression or the
extended scaling expression.  The two scalings are broadly equivalent
in that data for all sizes are included in the scaling
plots. Depending on the circumstances one or other scaling can be
easier to "read".

For specific normalized observables,
\begin{itemize}
  \item[-] when $Q$ is the normalized correlation length
    $\xi(\tau,L)T^{1/2}$, with $q=1$, 
    \begin{equation}
      Q(\tau,L)/(LT^{1/2})^{q} = \xi(\tau,L)/L
    \end{equation}
    so the scaling rule is~\cite{campbell:06}
    \begin{equation}
      \xi(\tau,L)/L = F^*[(LT^{1/2})^{1/\nu}\tau]
    \end{equation}
  \item[-] when $Q$ is $\chi(\tau,L)$, with $q = 2-\eta$, the scaling
    rule is~\cite{campbell:06}
    \begin{equation}
      \chi(\tau,L)/(LT^(1/2))^{2-\eta} = F^*[(LT^{1/2})^{1/\nu}\tau]
    \end{equation}
  \item[-] for an Ising model when $Q$ is the normalized Binder
    cumulant $L^{D}g(\tau,L)$, with $q = D$ if hyperscaling holds,
    \begin{equation}
      \frac{Q(\tau,L)}{(LT^{1/2})^{q}} = \frac{L^{D}g(\tau,L)}{(LT^{1/2})^{D}} =
      \frac{g(\tau,L)}{T^{D/2}}
    \end{equation}
    so the scaling rule is
    \begin{equation}
      g(\tau,L)/T^{D/2} = F^*[(LT)^{1/\nu}\tau]
    \end{equation}
    (This expression was not cited in Ref.~\cite{campbell:06}).  For
    the particular case of the $D=5$ Ising model, $\nu = 1/2, \eta =
    0$ and the standard hyperscaling rules do not hold. The modified
    scaling rules are discussed below in Section~\ref{sec:VI}.
\end{itemize}

Again, for an ISG $T^{1/2}$ is replaced throughout by $T$.  Finally,
as the scaling rules for the correlation length and for the Binder
cumulant have the same $x$-axis $(LT^{1/2})^{1/\nu}\tau$, if
hyperscaling holds a further scaling plot is given by $y(T,L) =
g(T,L)/T^{D/2}$ against $x(T,L)=\xi(T,L)/L$ (or for an ISG $y(T,L) =
g(T,L)/T^{D}$ against $x(T,L)=\xi(T,L)/L$). This is a particularly
remarkable format as both $x(T,L)$ and $y(T,L)$ represent purely
measured data sets; no inputs concerning the values of $T_c$ or of the
critical exponents are required for the scaling. This type of plot can
represent a stringent test of hyperscaling.

\section{The 3D Ising model}\label{sec:V}
\begin{figure}
  \includegraphics[width=3.4in]{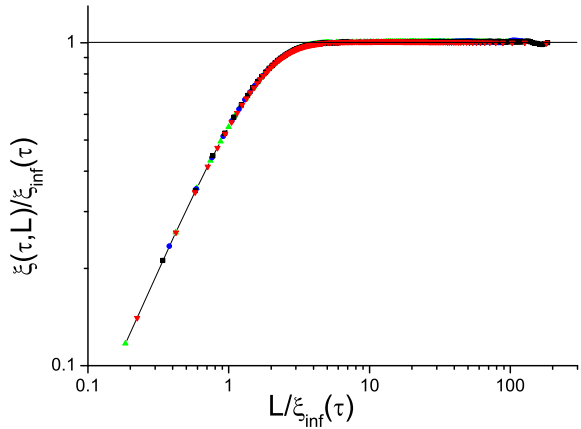}
  \caption{(Color on line) Dimension 3 Ising model Privman-Fisher
    finite-size scaling plot for the correlation length. Sample sizes
    $L=32$ (black squares), $L=24$ (blue circles), $L= 16$ (green
    triangles), $L=8$ (red inverted triangles). $\xi_{\infty}(\tau)=
    1.074\tau^{-0.630}\beta^{1/2}(1-0.12\tau^{0.5}+0.05\tau)$ }
  \protect\label{fig1}
\end{figure}

\begin{figure}
  \includegraphics[width=3.4in]{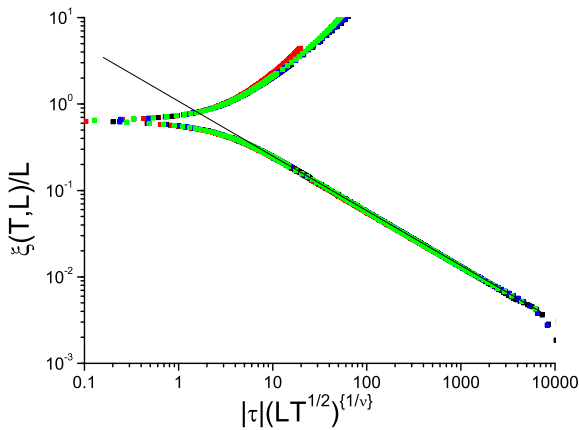}
  \caption{(Color on line) Dimension 3 Ising model extended scaling
    plot for the correlation length. Sample sizes as indicated in
    Fig.~\ref{fig1}. Upper branch $\beta >\beta_{c}$, lower branch
    $\beta < \beta_{c}$. The black straight line has slope $-\nu$.}
  \protect\label{fig2}
\end{figure}

We will first consider the canonical cubic lattice Ising model in
dimension $3$, in order to exhibit a case where hyperscaling is well
established.  Critical temperature $\beta_{c}$ and the critical
exponents are known to very high precision in this model
\cite{simmons:15}.  Simulation data were mainly taken from
Ref.~\cite{haggkvist:07}. See Ref.~\cite{campbell:11} for detailed
Privman-Fisher and extended scaling analyses of $\chi(\tau,L)$ in this
model including the correction terms. Scalings for the correlation
length $\xi(\tau,L)$ are shown in Figs.~\ref{fig1} and \ref{fig2}.

The ThL susceptibility and normalized correlation length corrections
are relatively small for both observables \cite{campbell:11} and the
extended scaling expressions with only two leading Wegner correction
terms give rather accurate fits to the ThL simulation and HTSE data
over the whole paramagnetic temperature range.
(Scaling with $\xi(t,L)$ rather than with $\xi(\tau,L)/\beta^{1/2}$
leads to a high temperature "cross-over" behavior in Ising
ferromagnets \cite{luijten:97} which is an artefact \cite{lundow:11}).

\begin{figure}
  \includegraphics[width=3.4in]{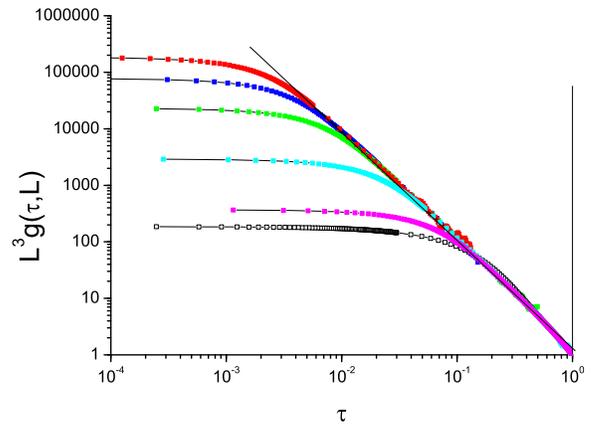}
  \caption{(Color on line) Dimension 3 Ising model normalized Binder
    cumulant $L^{3}g(\tau,L)$ as a function of $\tau$ for $L=64$,
    $48$, $32$, $16$, $8$ (top to bottom). The open points are the sum
    of the first 8 HTSE values evaluated from
    Ref.~\cite{butera:02a}. Exponent $1.89=3\nu$; fit
    $(L^{3}g(\tau,L))_{\infty} = 1.57\tau^{-1.89}(1 - 0.294\tau -
    0.069\tau^{2.85})$. Slope $-1.89$ black curve.}
  \protect\label{fig3}
\end{figure}

\begin{figure}
  \includegraphics[width=3.4in]{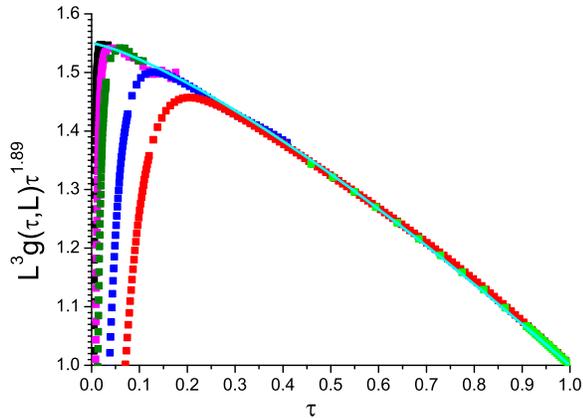}
  \caption{(Color on line) Dimension 3 Ising model normalized Binder
    cumulant $L^3g(\tau,L)\tau^{1.89}$ correction factor as a function
    of $\tau$ for $L=48$, $32$, $24$, $12$, $8$ (top to bottom). Fit
    of $L^3g(\tau,L)\tau^{1.89}=1.55(1 - 0.294\tau -
    0.069\tau^{2.85})$ cyan curve.}  \protect\label{fig4}
\end{figure}

For the Binder cumulant in the $3$D Ising model, ThL simulation and
HTSE data (evaluated from the tabulated series in Ref.~\cite
{butera:02a}) can be fitted satisfactorily by
\begin{equation}
  L^D g(\tau,L) = 1.57\tau^{-1.89}\left(1 - 0.294\tau - 0.069\tau^{2.85}\right)
  \label{3Dgscaled}
\end{equation}
where the critical exponent is equal to $D\nu= 3\cdot 0.63$ as
expected from hyperscaling, see Figs.~\ref{fig3} and \ref{fig4}.  The
amplitude of the expected leading confluent correction term
proportional to $\tau^{0.523}$ turns out to be negligible for the
Binder cumulant in the 3D Ising universality class \cite{lundow}; the
next effective correction terms proportional to $\tau$ and to
$\tau^{\theta_{\mathrm{eff}}} \sim \tau^{2.85}$ dominate the
corrections

\begin{figure}
  \includegraphics[width=3.4in]{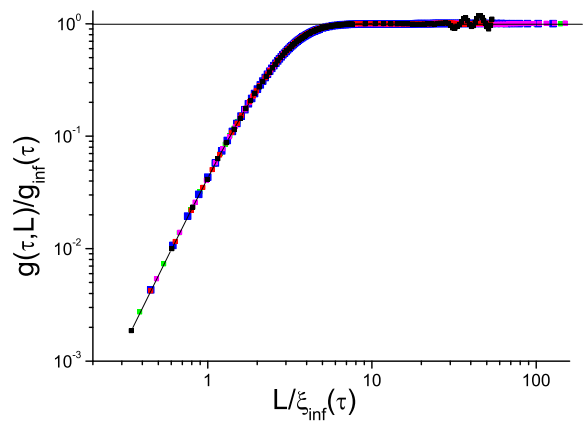}
  \caption{(Color on line) Dimension 3 Ising model Privman-Fisher plot
    of the Binder cumulant. Point coding as in Fig.~\ref{fig4}; $\nu =
    0.63$. The $g_{\infty}(\tau)$ as in caption of Fig.~\ref{fig3} and
    $\xi_{\infty}(\tau)$ as in caption of Fig.~\ref{fig1}.}
  \protect\label{fig5}
\end{figure}

\begin{figure}
  \includegraphics[width=3.4in]{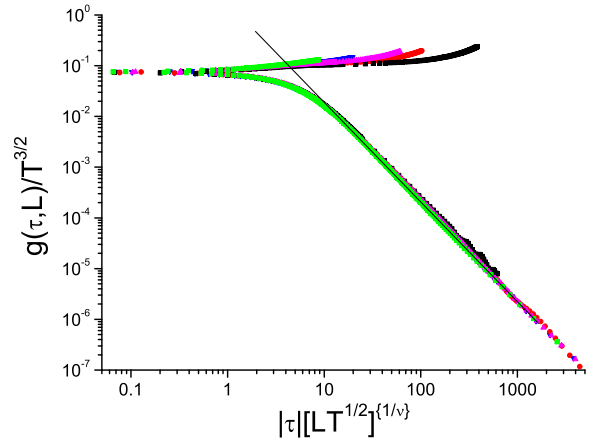}
  \caption{(Color on line) Dimension 3 Ising model extended scaling
    plot of the Binder cumulant. Point coding as in
    Fig.~\ref{fig4}. The straight line has slope $-3\nu$, with $\nu =
    0.63$. } \protect\label{fig6}
\end{figure}

Privman-Fisher and extended scaling format plots of the normalized
Binder cumulant $L^{3}g(\tau,L)$ for all measured $L$ and $\tau$ are
shown in Figs.~\ref{fig5} and \ref{fig6}.  The overall scalings hold
well for all temperatures and for all sizes from infinity down to
criticality (and even somewhat beyond as the extended scaling plots
show), as to be expected for a model where hyperscaling holds.

\begin{figure}
  \includegraphics[width=3.4in]{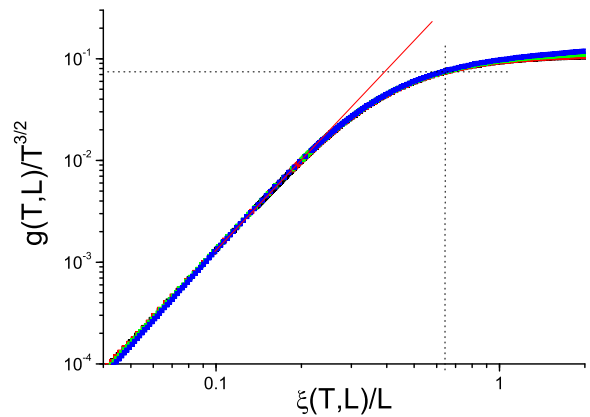}
  \caption{(Color on line) Dimension 3 Ising model normalized scaling
    plot of Binder cumulant against correlation length with no input
    parameters. Sizes shown are $L=32$ (black squares), $L=24$ (red
    circles), $L=12$ (green triangles), $L=8$ (blue inverted
    triangles). Dashed lines indicate criticality. }
  \protect\label{fig7}
\end{figure}

As a direct test of the Binder cumulant hyperscaling we can make up a
pure data-against-data plot of $y(T,L) = g(T,L)/T^{3/2}$ against
$x(T,L)= \xi(T,L)/L$, see Fig.~\ref{fig7}, again assuming hyperscaling
so $q=3$.  The overall scaling is good over the whole temperature
range shown. (The empirical $g(T,L)$ against $\xi(T,L)$ form of
scaling of the same data, which has been suggested for instance in
Ref.~\cite{jorg:06}, appears satisfactory when presented as a
linear-linear plot but is unsatisfactory when presented as a log-log
plot covering the entire temperature range).

\section{The 5D Ising model}\label{sec:VI}
\begin{figure}
  \includegraphics[width=3.4in]{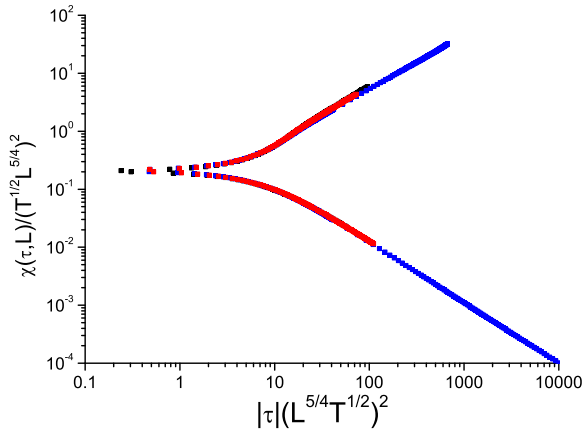}
  \caption{(Color on line) Dimension 5 Ising model. Susceptibility
    extended scaling. Sizes shown are $L=24$ (black squares), $L=16$
    (red circles), $L=8$ (blue triangles). Upper branch $\beta >
    \beta_{c}$.}  \protect\label{fig8}
\end{figure}

The upper critical dimension of standard Ising models is
$D_{\mathrm{ucd}} = 4$. For higher dimensions critical exponents are
mean field (MF) and independent of $D$ : $\alpha=0$, $\nu =1/2$,
$\gamma=1$, $\eta = 0$, $\omega=1$. It is well known that the standard
hyperscaling relation $\alpha = 2 - D\nu$ cannot hold above $D = 4$ as
this relation is incompatible with the MF exponents.  A general rule
for the FSS correlation length above $D_{\mathrm{ucd}}$ which holds
both for periodic boundary conditions and for free boundary
conditions, is $\xi(\tau,L) \sim L^{\koppa}$ with
$\koppa=D/D_{\mathrm{ucd}}$ \cite{berche:12}. Indeed it has been
shown that above $D=4$ at criticality the effective finite-size
correlation length $\xi(\tau_{c},L)$ scales with $L^{D/4} =
L^{\koppa} = L^{D/D_{\mathrm{ucd}}}$ rather than with $L$ while
the critical Binder cumulant $g(\tau_{c},L)$ remains independent of
$L$ (to within corrections to scaling)~\cite{jones:11,berche:12}.  In
the Privman-Fisher and extended scaling plots shown in
Figs.~\ref{fig8}, \ref{fig10} and \ref{fig11} $L^{D/D_{\mathrm{ucd}}}$
replaces $L$ everywhere, showing that the FSS rule for $D$ above
$D_{\mathrm{ucd}}$ is valid in the whole paramagnetic temperature
range.

\begin{figure}
  \includegraphics[width=3.4in]{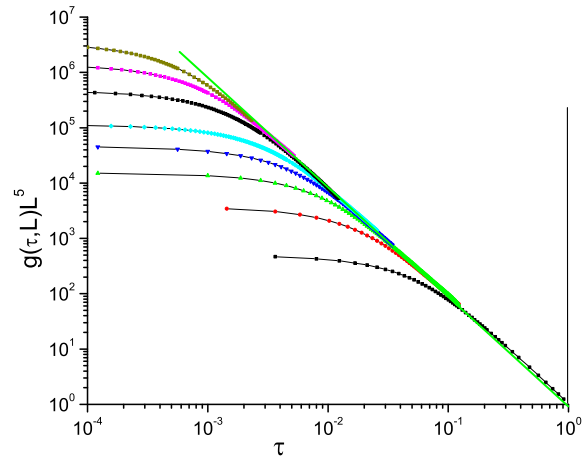}
  \caption{(Color on line) Dimension 5 Ising model. Normalized Binder
    cumulant against $\tau$ for $L = 24$, $20$, $16$, $12$, $10$, $8$,
    $6$, $4$ (top to bottom). Fit : full green curve.}
  \protect\label{fig9}
\end{figure}

\begin{figure}
  \includegraphics[width=3.4in]{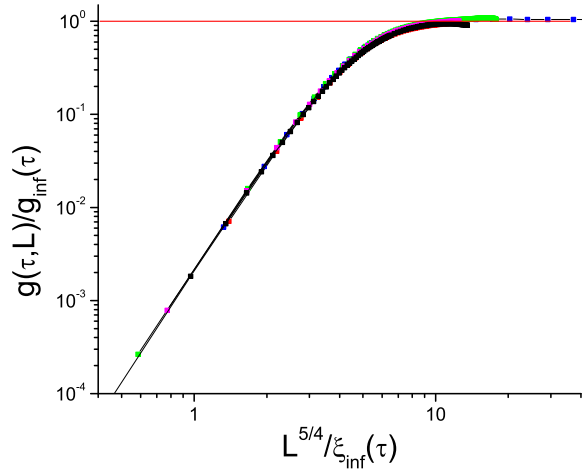}
  \caption{(Color on line) Dimension 5 Ising model Binder cumulant
    Privman-Fisher scaling for $L=16$ (black squares), $L=10$ (pink
    circles), $L=8$ (green triangles), $L=6$ (red inverted triangles),
    $L=4$ (blue diamonds).} \protect\label{fig10}
\end{figure}

\begin{figure}
  \includegraphics[width=3.4in]{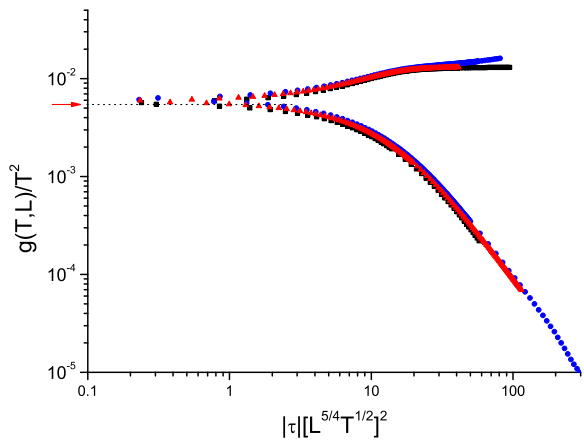}
  \caption{(Color on line) Dimension 5 Ising model extended scaling of
    the Binder cumulant for $L=24$ (black squares), $L=12$ (red
    triangles), $L=6$ (blue circles). Upper branch $\beta >
    \beta_{c}$. Arrow : critical value from
    Ref.~\cite{jones:11}.}\protect\label{fig11}
\end{figure}

The MF value of the gap exponent is $\Delta = 3/2$ \cite{butera:12},
so the second standard hyperscaling rule $2\Delta = D\nu +\gamma$ must
also be violated above $D = 4$. From the same reasoning as above, it
follows that this hyperscaling breakdown leads to a MF ThL exponent
for the normalized Binder cumulant which is $2$ rather than $D\nu =
D/2$, so in the ThL $L^{D}g(\tau,L) \sim \tau^{-2}$. Simulation data
in dimension $5$ for $L^{5}g(\tau,L)$ as a function of $\tau$ are
shown in Fig.~\ref{fig9} where it can be seen that in the entire ThL
regime from infinite temperature to criticality this rule indeed holds
with a critical exponent $2$ and a weak correction to scaling. The
critical amplitudes for $\chi_{4}$ and $\chi$ are $\sim 1.40$
\cite{butera:12a} and $\sim 1.29$ \cite{berche:08,butera:12a}
respectively, and the leading thermal correction exponent is $\theta =
\omega\nu = 1/2$, so including the leading correction term the ThL
scaling is $L^{D}g(\tau,L) = 0.80\tau^{-2}(1+0.176\tau^{1/2})$ (It
should be noticed that for $D=5$ the form of the normalization of the
Binder cumulant remains $L^{5}g(\tau,L)$, and does not become
$L^{25/4}g(\tau,L)$ because $L^5$ is just the number of spins).

This behavior would not have been recognized easily if the
conventional reduced temperature $t= 1-T/T_{c}$ had been used as the
scaling parameter.  The Privman-Fisher Binder cumulant scaling
(without correction terms) with $\nu = 1/2$, $q = 4$, $L^{5}g(\tau,L)
\sim \tau^{-2}$, $\xi(\tau,\infty) = \beta^{1/2}\tau^{-1/2}$ and
$L^{5}g(\tau,L)/L^{5}g(\tau,\infty) = F[L^{5/4}/\xi(\tau,\infty)]$
becomes $g(\tau,L)/T^{2} = F^{*}[L^{5/2}\tau T]$ which is consistent
with an $L$ independent $g(0,L) = F^{*}(0)$ at criticality ($\tau =
0$, $T = T_{c}$) to within correction terms.

Correction terms can be included in the Privman-Fisher scaling. We
have no simulation or HTSE data for the correlation length in
dimension $5$. However, we assume that the ThL correlation length
behaves as $\xi(\tau,\infty) =
C\beta^{1/2}\tau^{-1/2}[1-(1-1/C)\tau^{1/2}]$ with a weak leading
order scaling correction (as observed for Ising models in dimensions
$2$ and $3$ \cite{campbell:08,campbell:11}).  The dimension $5$
modified Privman-Fisher scaling rule for an observable $Q$ is
\begin{eqnarray}
  \frac{Q(\tau,L)}{Q(\tau,\infty)} = F[L^{5/4}/\xi(\tau,\infty)]\nonumber \\ =
  F[L^{5/4}/(C\beta^{1/2}\tau^{-1/2}(1-(1-1/C)\tau^{1/2}))]
\end{eqnarray}

Scaled data for for the normalized Binder cumulant $Q =
L^{5}g(\tau,L)$ and $q_{g} = 2/\nu= 4$, are shown in Fig.~\ref{fig10}
and Fig.~\ref{fig11}. With the correlation length critical amplitude
chosen as $C = 0.75$ to optimize the $\chi$ scaling, both scalings are
excellent. This validates the correlation length normalization form
$\xi(\tau,\infty)/\beta^{1/2}$ in 5D, and the replacement of $L$ by
$L^{5/4}$ in the Privman-Fisher correlation length scaling rule not
only in a narrow critical region but in the entire paramagnetic
regime.  The plots cover data for all sizes from infinite temperature
to criticality (the left hand $\tau=0$ limit in Fig.~\ref{fig10}) and
even to well beyond criticality (the upper branch in
Fig.~\ref{fig11}).

In the FSS limit close to criticality this scaling implies that
$g(\tau_{c},L)$ is independent of $L$ (to within corrections to
scaling), which is consistent with the data of
Ref.~\cite{jones:11}. The preservation of the rule of size
independence for the dimensionless Binder cumulant at criticality
results from the combined effects of the two hyperscaling
breakdowns. The susceptibility $\chi(\tau,L)$ finite-size scaling
becomes $\chi(\tau_{c},L) \sim L^{5/2}$ at criticality. If data were
available, the overall Privman-Fisher correlation length scaling rule
would be
\begin{equation}
  \xi(\tau,L)/L^{5/4} = F[L^{5/4}/(\beta^{1/2}\tau^{-1/2})]
 \end{equation}
so with $\xi(\tau_{c},L)/L^{5/4}$ independent of $L$ at criticality to
within the correction term, as observed in Ref.~\cite{jones:11}.  An
analysis of Ising data in dimension $6$ shows that they follow just
the same rules as in dimension $5$ {\it mutatis mutandis} (so with
$\koppa =3/2$), again over the entire paramagnetic regime.  Thus the
ThL $L^6g(\tau, L)$ has critical exponent $2$ as in 5D, and the
Privman Fisher scaling rules all work with $x$-axis
$L^{6/4}/\xi(\tau,\infty)$ for all temperatures.

\section{Ising spin glasses}\label{sec:VII}
Now we turn to ISGs. The standard ISG Hamiltonian is $\mathcal{H}= -
\sum_{ij}J_{ij}S_{i}S_{j}$ with the near neighbor symmetric random
distributions normalized to $\langle J_{ij}^2\rangle=1$. The
normalized inverse temperature is $\beta = (\langle
J_{ij}^2\rangle/T^2)^{1/2}$. The Ising spins live on simple
hyper-cubic lattices with periodic boundary conditions. For the
bimodal models $J_{ij} =\pm 1$ at random. The spin overlap parameter
is defined as usual by
\begin{equation}
  q = \frac{1}{L^{D}}\sum_{i} S^{A}_{i}S^{B}_{i}
\end{equation}
where $A$ and $B$ indicate two copies of the same system.  Klein {\it
  et al.}~\cite{klein:91} quote exactly the same hyperscaling relation
Eq.~\eqref{gam4} for $\chi_{4}$ in the ISGs as in the Ising
ferromagnets (with the spin overlap moments $\langle q^2\rangle$ and
$\langle q^4\rangle$ replacing the magnetization moments $\langle
m^2\rangle$ and $\langle m^4\rangle$), so the RGT hyperscaling
prediction for the ISG Binder cumulant critical exponent is again
$\gamma_{4}- 2\gamma = D\nu$. Because the interaction parameter in the
ISGs is $\langle J_{ij}^2\rangle$ the appropriate ISG temperature
scaling variable is $\tau = 1-(\beta/\beta_{c})^2$
\cite{daboul:04,campbell:06} and the appropriate normalized
correlation length is $\xi(\tau,L)/\beta$ \cite{campbell:06}.

Some of the simulation data in the ISGs are the same as those in
Refs.~\cite{lundow:15,lundow:15a} where the simulation techniques have
already been described in detail. Means were taken on at least $8192$
samples for each $L$ with of the order of $40$ different
temperatures. The maximum size studied was $L = 32$.
Particular attention was paid to achieving full equilibration.  For
the $3$D bimodal model comparisons with tabulated data generously
provided by H.~Katzgraber and by K.~Hukushima from independent
simulations, and from raw data tabulations related to
Ref.~\cite{hasenbusch:08} helpfully published on line by Hasenbusch,
Pelissetto and Vicari, confirm equilibration. Unfortunately the
maximum sizes $L$ for simulations in ISGs have been limited in
practice by the available computational ressources and we know of no
published measurements to $L$ greater than $40$ in $D=3$.  For the 3D
bimodal ISG extended scaling plots for $\chi(\tau,L)$ and
$\xi(\tau,L)/\beta$ were shown in Ref.~\cite{campbell:06} with fit
values for the critical parameters close to those estimated later from
FSS analyses \cite{katzgraber:06,hasenbusch:08,baity:13}.  Assuming
$\beta_{c}=0.9075$ (from Ref.~\cite{baity:13}, which is consistent with
the estimates of Refs.~\cite{katzgraber:06} and \cite{hasenbusch:08}
to within the quoted errors) we have made plots of the temperature
dependent effective exponents $\gamma(\tau,L)$ and $\nu(\tau,L)$
\cite{lundow}.

\begin{figure}
  \includegraphics[width=3.4in]{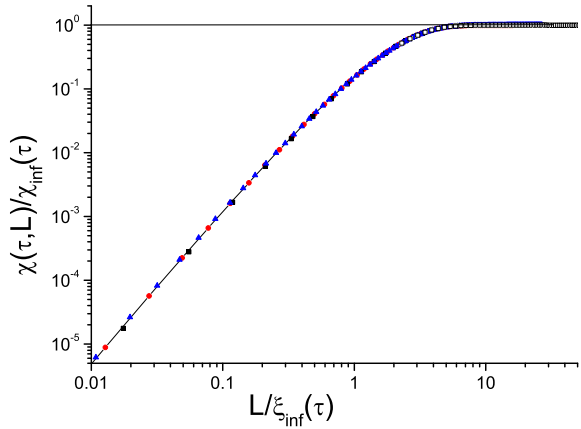}
  \caption{(Color on line) Dimension 3 bimodal interaction ISG
    Privman-Fisher susceptibility scaling for $L=32$ (open squares),
    $L=20$ (black squares), $L=10$ (red circles), $L=6$ (blue
    triangles).  For $\chi_{\infty}(\tau), \xi_{\infty}(\tau)$, see
    text. } \protect\label{fig12}
\end{figure}

\begin{figure}
  \includegraphics[width=3.4in]{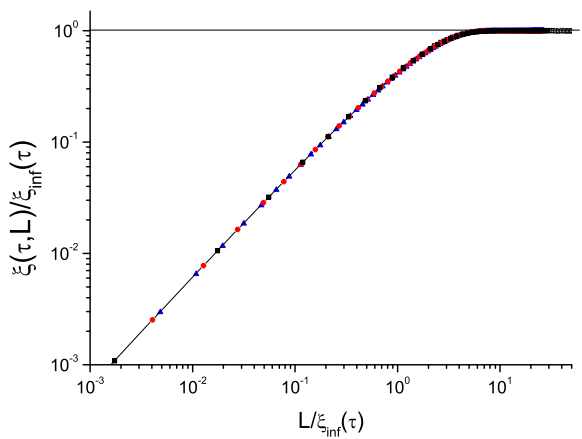}
  \caption{(Color on line) Dimension 3 bimodal interaction ISG
    Privman-Fisher correlation length scaling for $L=32$ (open
    squares), $L=20$ (black squares), $L=10$ (red circles), $L=6$
    (blue triangles). For $\xi_{\infty}(\tau)$, see text.}
  \protect\label{fig13}
\end{figure}

Extrapolating to criticality, the estimations for the critical
exponents are $\gamma = 6.15(10)$ in agreement with
Ref.~\cite{baity:13}, and $\nu = 2.45(10)$, in good agreement with the
estimate of Ref.~\cite{hasenbusch:08} (but significantly lower than
the estimate of Ref.~\cite{baity:13}). With these values in hand we
fit the ThL susceptibility and correlation length data with leading
effective corrections \cite{lundow} 
\begin{eqnarray}
  \chi(\tau) &=&
  1.28\tau^{-6.15}\left(1-0.62\tau^{2.45}+0.405\tau^{7}\right)\\ \xi(\tau)
  &=& 1.15\tau^{2.45}\beta\left(1-0.24\tau^{2.45}+0.11\tau^{7}\right)
\end{eqnarray} 
It can be noted that when the $\beta$ normalization factor is
included, the correction terms for the normalized $\xi(\tau)/\beta$
are weak over the entire temperature range, as already observed in
\cite{campbell:06}. (In the dimension 3 bimodal ISG for both
observables it is essential to introduce two correction terms for a
satisfactory fit over the whole paramagnetic temperature range. This
implies that in principle two corresponding correction terms should
also be included in FSS analyses).  Standard Privman-Fisher scaling
for $\chi(\tau,L)$ and for the normalized correlation length
$\xi(\tau,L)/\beta$ with the correction terms are shown in
Figs.~\ref{fig12} and \ref{fig13}, now including all the data and not
just the ThL data. The overall scaling for all paramagnetic
temperatures and all sizes is excellent which in particular is
consistent with the extrapolations to criticality being valid.

\begin{figure}
  \includegraphics[width=3.4in]{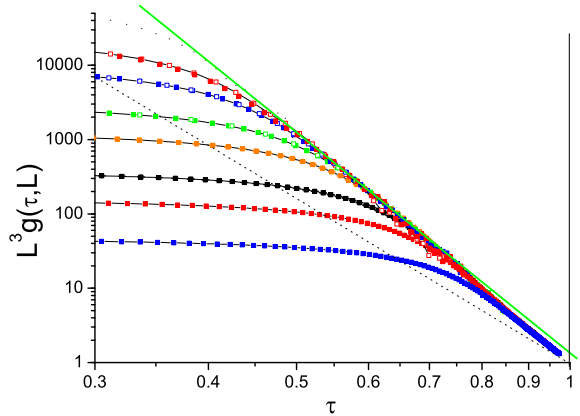}
  \caption{(Color on line) Dimension 3 bimodal interaction ISG
    normalized Binder cumulant against $\tau$ for $L= 32$, $24$, $16$,
    $12$, $8$, $6$, $4$ (top to bottom). Dashed line : slope $7.35 =
    3\nu$, full green line slope $10.3$. } \protect\label{fig14}
\end{figure}

In Fig.~\ref{fig14} the normalized Binder cumulant $L^{3}g(\tau,L)$
against $\tau$ data are shown.  Sizes $L$ are limited by the rapidly
increasing numbers of spins and by equilibration difficulties; the
lowest value of $\tau$ where the ThL condition still holds is only
about $0.5$.  Down to this lowest accessible ThL value of $\tau$, the
ThL data can be fitted approximately by $L^{3}g(\tau,L) \sim
\tau^{-10.5}$ with a correction term, so an effective exponent much
larger than the hyperscaling value $3\nu = 7.35$. If data for much
higher $L$ (and so to lower $\tau$-values still within the ThL regime)
were available there seems no {\it a priori} reason to expect this
behavior with a large effective exponent to change.

\begin{figure}
  \includegraphics[width=3.4in]{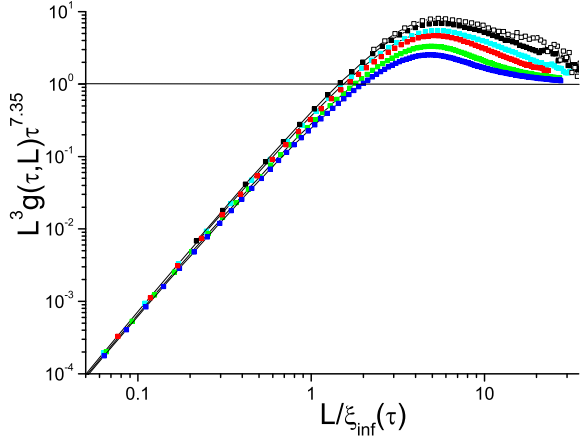}
  \caption{(Color on line) Dimension 3 bimodal interaction ISG
    normalized Binder cumulant Privman-Fisher plot assuming
    hyperscaling. Sizes shown are $L= 32$, $24$, $16$, $12$, $8$, $6$
    (top to bottom).}  \protect\label{fig15}
\end{figure}

\begin{figure}
  \includegraphics[width=3.4in]{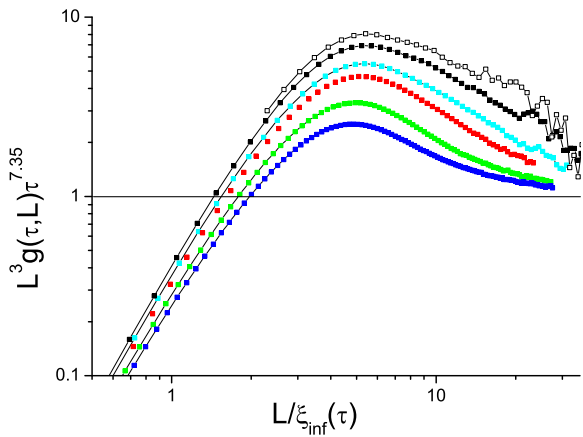}
  \caption{(Color on line) Dimension 3 bimodal interaction ISG
    normalized Binder cumulant Privman-Fisher plot assuming
    hyperscaling. Zoom of Fig.~\ref{fig15}.} \protect\label{fig16}
\end{figure}

In Figs.~\ref{fig15} and \ref{fig16} the Privman-Fisher scaling curves
are shown for the same 3D bimodal normalized Binder cumulant data
assuming hyperscaling, i.e. with a critical exponent equal to $7.35$.
In the ThL regime (on the right) the scaled curves have peaks
increasing dramatically in size, and moving to the left regularly with
increasing $L$. The peaks reach values of the order of $10$ for the
sizes covered by the present measurements, far from remaining near
$y(x) \sim 1$ as would be expected if the hyperscaling rule was obeyed
and as is observed above for $\chi(\tau,L)$ and $\xi(\tau,L)/\beta$
Privman-Fisher plots where hyperscaling is not involved.  (While
scaling analyses of the $\chi(\tau,L)$ and $\xi(\tau,L)T$ data for the
3D bimodal and Gaussian ISGs show typical critical amplitudes $C \sim
1.10$ associated with weak correction terms~\cite{lundow}, attempts to
fit the $L^{3}g(\tau,L)$ data with critical exponent $D\nu$ plus
corrections lead to huge critical amplitudes $C \sim 25$, and exotic
correction terms. This interpretation appears unphysical).
However, on leaving the ThL regime as temperatures tend towards
criticality, the hyperscaling behavior with curves for all $L$
overlapping is gradually recovered in the FSS regime.

\begin{figure}
  \includegraphics[width=3.4in]{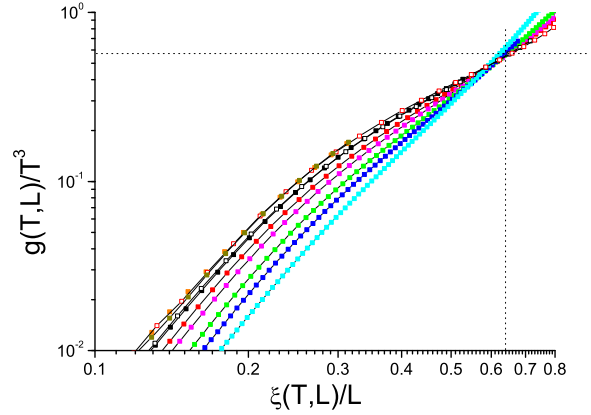}
  \caption{(Color on line) Dimension 3 bimodal interaction
    ISG. Normalized Binder cumulant scaling of $g(\tau,L)/T^3$ against
    $\xi(\tau,L)/L$ assuming hyperscaling} \protect\label{fig17}
\end{figure}

An extended scaling plot assuming hyperscaling, $g(\tau,L)/T^3$
against $\xi(\tau,L)/L$ in Fig.~\ref{fig17} should show curves for all
$L$ overlapping as for the 3D Ising model, see Fig.~\ref{fig7}. Again
there are strong deviations in the ThL regime on the left (and also
for $\beta > \beta_{c}$ on the right) with hyperscaling behavior
restored at criticality indicated by the dashed lines.

\begin{figure}
  \includegraphics[width=3.4in]{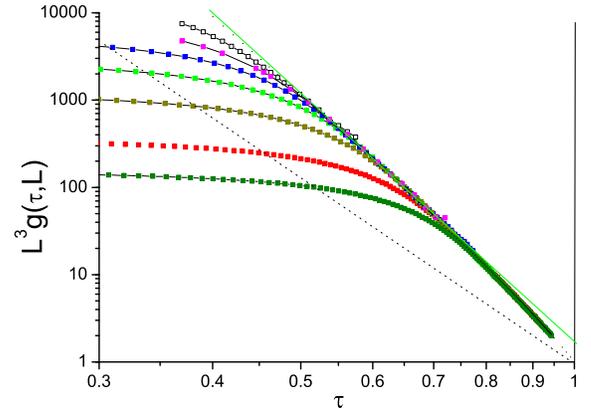}
  \caption{(Color on line) Dimension 3 Gaussian interaction ISG
    normalized Binder cumulant against $\tau$ for $L= 32$, $24$, $20$,
    $16$, $12$, $8$, $6$ (top to bottom). Dashed line : slope $7.15 =
    3\nu$, full green line slope $9.5$. } \protect\label{fig18}
\end{figure}

\begin{figure}
  \includegraphics[width=3.4in]{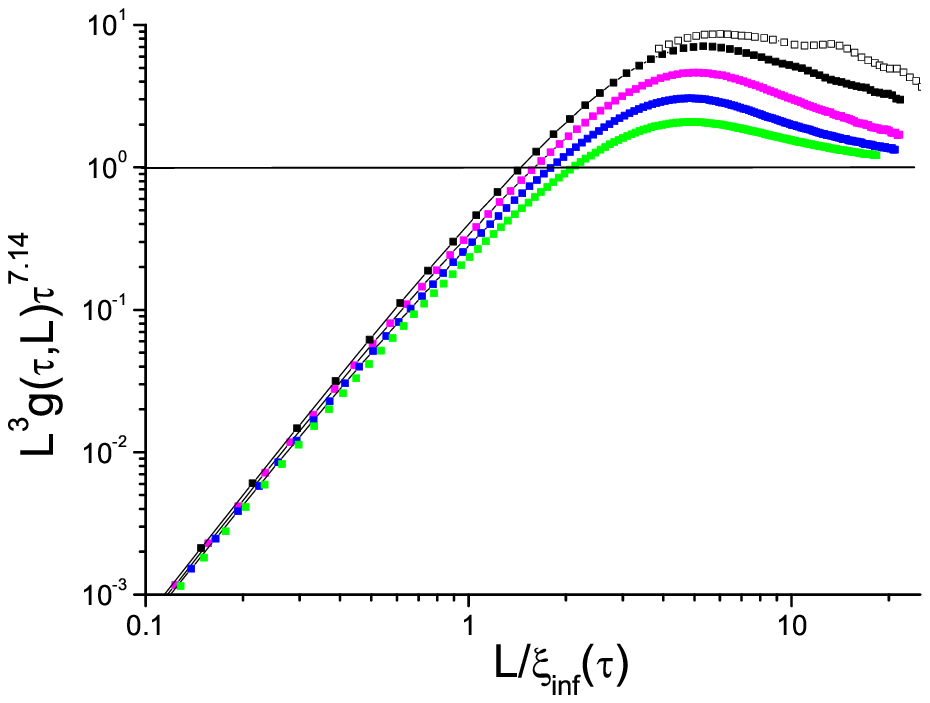}
  \caption{(Color on line) Dimension 3 Gaussian interaction ISG
    normalized Binder cumulant Privman-Fisher scaling assuming
    hyperscaling. Sizes shown are $L= 32$, $20$, $10$, $6$, $4$ (top
    to bottom).}  \protect\label{fig19}
\end{figure}

\begin{figure}
  \includegraphics[width=3.4in]{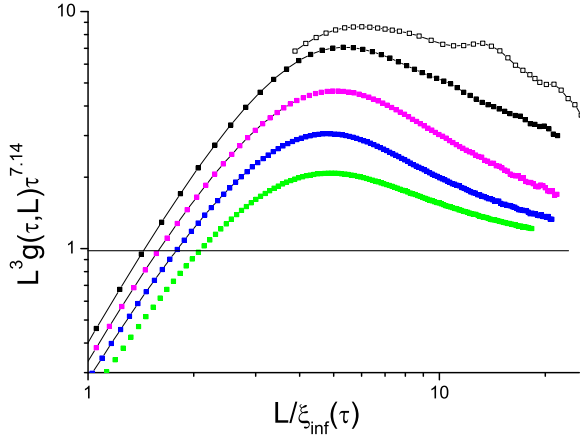}
  \caption{(Color on line) Dimension 3 Gaussian interaction ISG
    normalized Binder cumulant Privman-Fisher scaling assuming
    hyperscaling. Sizes shown are $L= 32$, $20$, $10$, $6$, $4$ (top
    to bottom). Zoom of Fig.~\ref{fig19}.} \protect\label{fig20}
\end{figure}

The most recent estimates of the dimension 3 Gaussian interaction ISG
critical parameters \cite{katzgraber:06} are $\beta_c = 1.05(1)$, $\nu
= 2.44(9)$, and $\eta = -0.37(5)$. The Gaussian model shows
qualitatively very much the same Binder cumulant behavior as the
bimodal model, see Figs.~\ref{fig18}, \ref{fig19} and \ref{fig20}. The
data show $L^3g(\tau,L) \sim \tau^{-10.0}$ in the ThL regime, an
effective exponent which is much larger than the hyperscaling value
$3\nu \sim 7.2$. Again in the Privman-Fisher plots there are strong
peaks in the ThL regime rather than the horizontal line $y(x) \sim 1$
expected on the hyperscaling assumption. As for the bimodal model
there is a return to hyperscaling in the FSS regime.

\begin{figure}
   \includegraphics[width=3.4in]{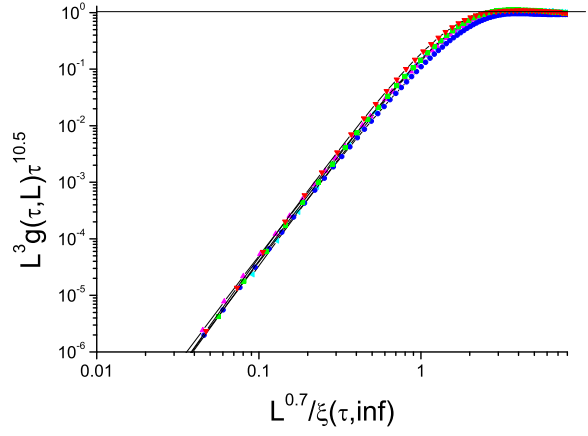}
   \caption{(Color on line) Dimension 3 bimodal ISG model. Normalized
     Binder cumulant $L^{3}g\tau^{10.5}$ against $L^{0.7}/\xi(\tau)$
     for $L = 24$ (cyan left triangle), $L=16$ (red inverted
     triangle), $L=12$ (green square), $L=8$ (pink triangle), $L=6$
     (blue circle). } \protect\label{fig21}
\end{figure}

\begin{figure}
  \includegraphics[width=3.4in]{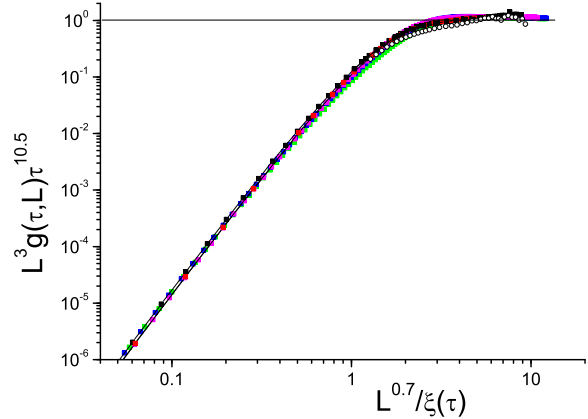}
  \caption{(Color on line) Dimension 3 Gaussian ISG model. Normalized
    Binder cumulant $L^{3}g\tau^{10.5}$ against $L^{0.7}/\xi(\tau)$
    for $L = 32$ (open circle), $L=24$ (cyan left triangle), $L= 16$
    (red inverted triangle), $L=10$ (pink triangle), $L=6$ (blue
    circle), $L=4$ (green square).}  \protect\label{fig22}
\end{figure}

There is a possible empirical rationalization of the ISG Binder
cumulant behavior.  The data show that the ISG correlation length
follows the ThL rule $\xi(\tau) = T\tau^{-2.4}$ (so $\nu(\tau) \sim
2.4$) to a good approximation over the entire paramagnetic range of
temperatures.  If we assume that the observed ThL behavior
$L^{3}g(\tau) \sim \tau^{-10.5}$ extends to criticality also, it is
equivalent to assuming that the strong disorder modifies the rule
$L^{3}g(\tau) \sim \tau^{-\nu D}$ to $L^{3}g(\tau) \sim \tau^{-\nu
  D_{\mathrm{eff}}}$ with $D_{\mathrm{eff}} \sim 4.5$.  By analogy
with the Ising dimension 5 formalism we can write $D_{\mathrm{eff}}
=3/\koppa$ with $\koppa \sim 0.7$ for both bimodal and Gaussian ISGs,
and the modified Privman-Fisher scalings of the normalized Binder
cumulants (see Fig.~\ref{fig10} for 5D Ising) should then take the
form $L^{3}g(\tau,L)/(\tau^{-3\nu/\koppa})$ against
$L^{\koppa}/(T\tau^{-\nu})$.  These plots are shown in
Figs.~\ref{fig21} and \ref{fig22} for the 3D bimodal and Gaussian ISGs
respectively.

\section{Conclusion}\label{sec:VIII}

The scaling of the susceptibility, the normalized correlation length,
and the normalized Binder parameter are discussed for Ising models in
dimensions 3 and 5, and for ISG models in dimension 3, using a
rational normalization and scaling approach which covers the entire
paramagnetic temperature region and not just the finite-size scaling
regime.

For the canonical dimension 3 Ising model, the observed scaling of the
normalized Binder cumulant $L^{D}g(\tau,L)$ is fully consistent with
hyperscaling over the entire temperature range as to be expected. For
the dimension 5 Ising model, above the upper critical dimension, the
susceptibility, normalized correlation length, and normalized Binder
cumulant scaling, are consistent with mean field exponents and so with
the known breakdown of hyperscaling, over the entire temperature range
including both the thermodynamic limit and the finite-size scaling
regimes. The breakdowns of the two hyperscaling rules in dimension 5
conspire to ensure the size independence of the dimensionless Binder
cumulant $g(\tau,L)$ at criticality.

In Ising spin glasses in dimension 3 the normalized Binder cumulant
scaling shows a clear breakdown of the standard scaling rule in all
the thermodynamic limit regime attainable with the available
computational facilities; there is a return to behavior compatible
with hyperscaling in the finite-size scaling regime at the approach to
criticality. As no such breakdown is observed for the other
observables, $\chi(\tau,L)$ and $\xi(\tau,L)T$, where hyperscaling is
not involved, we propose that this behavior is the consequence of a
Schwartz hyperscaling breakdown. Fuller characterization would require
measurements to significantly higher sizes.

In the dimension 3 ISGs the ThL regime scaling is of the form
$L^{D}g(\tau,L) \sim \tau^{q}$ with an effective exponent $q$ which is
significantly higher than the hyperscaling value $q = D$. This scaling
rule, with hyperscaling breakdown for the Binder cumulant only,
implies $g(\tau_{c},L) \sim L^{D-q}$ at criticality and so is
incompatible with the $g(\tau_{c},L)$ independent of $L$ observed in
ISGs. We can postulate that in ISGs the fundamental physical rule
concerning the size independence at criticality of dimensionless
observables such as the Binder cumulant overrides the thermodynamic
limit scaling rule at and close to criticality.

\begin{acknowledgments}
  We would like to thank Professor A.~Aharony, Dr. P.~Butera and
  Dr. C.~M\"{u}ller for helpful comments, and H. Katzgraber and
  K. Hukushima for access to their raw numerical data.  The
  computations were performed on resources provided by the Swedish
  National Infrastructure for Computing (SNIC) at the High Performance
  Computing Center North (HPC2N) and Chalmers Centre for Computational
  Science and Engineering (C3SE).
\end{acknowledgments}


\end{document}